\documentclass[conference,letterpaper]{IEEEtran}
\IEEEoverridecommandlockouts

\usepackage[letterpaper,left=0.75in,top=1in,textwidth=7in,textheight=9in,columnsep=0.24in]{geometry}
\usepackage{cite}
\usepackage{amsmath,amssymb,amsfonts}
\usepackage[hyphens]{url}
\usepackage{graphicx}
\usepackage{etoolbox}
\usepackage{textcomp}
\usepackage{xcolor}
\usepackage{tikz}
\usepackage{stfloats}
\usepackage{algpseudocode}
\usepackage[most]{tcolorbox}
\usepackage[hidelinks]{hyperref}
\usepackage[nameinlink,noabbrev]{cleveref}

\newtcblisting{promptbox}[2][]{
  enhanced,
  breakable,
  colback=blue!4,
  colframe=blue!35!black,
  boxrule=0.5pt,
  arc=2mm,
  left=1.5mm,
  right=1.5mm,
  top=1mm,
  bottom=1mm,
  title={#2},
  fonttitle=\bfseries,
  listing only,
  listing options={
    basicstyle=\ttfamily\scriptsize,
    breaklines=true,
    breakatwhitespace=false,
    columns=fullflexible,
    keepspaces=true,
    showstringspaces=false
  },
  #1
}

\newcommand{\mytitle}{A Harness for Synthesizing Diverse Naturalistic Full-duplex Conversations}
\title{\mytitle}
\renewcommand{\footnotesize}{\fontsize{9}{11}\selectfont}
\makeatletter
\patchcmd{\@makecaption}{\scshape}{\upshape}{}{\PackageError{paper}{Cannot set table caption font}{Check the caption definition.}}
\newcommand{\name}[1]{\author{#1}}
\newcommand{\@address}{}
\newcommand{\address}[1]{\gdef\@address{#1}}
\renewcommand{\@maketitle}{%
  \newpage
  \vbox to 2.375in{%
    \vskip 0.375in
    \centering
    {\normalfont\fontsize{14}{17}\selectfont\bfseries\MakeUppercase{\@title}\par}%
    \nointerlineskip
    \vskip 0.2in
    {\normalfont\fontsize{9}{11}\selectfont\@author\par}%
    \vskip 6pt
    {\normalfont\fontsize{9}{11}\selectfont\@address\par}%
    \vfil
  }%
}
\renewcommand{\@IEEENORMtitlevspace}{\dimexpr0.1in-\topskip\relax}
\renewcommand{\@IEEEMINtitlevspace}{\dimexpr0.1in-\topskip\relax}

\makeatother

\name{
\begin{tabular}{@{}c@{\hspace{12pt}}c@{\hspace{12pt}}c@{\hspace{12pt}}c@{}}
Matthew Sun$^{1,*}$ & Vinay Kothapally$^{2}$ & Meng Yu$^{2}$ & Chao Huang$^{2}$ \\
matt.suncy@gmail.com & vkothapally@global.tencent.com & raymondmyu@global.tencent.com & chaochhuang@global.tencent.com
\end{tabular}\\[6pt]
\begin{tabular}{@{}c@{\hspace{18pt}}c@{\hspace{18pt}}c@{}}
Hao Zhang$^{3,\dagger}$ & Yixuan Zhang$^{2}$ & Steve Yves$^{2}$ \\
h.zhangnwpu@gmail.com & yixuazhang@global.tencent.com & steveyves@global.tencent.com
\end{tabular}
}
\address{
$^{1}$ Tencent Americas, Palo Alto, CA, USA \\
$^{2}$ Tencent Americas, Bellevue, WA, USA \\
$^{3}$ School of Electronic Information, Wuhan University, Wuhan, Hubei, China
}

\hypersetup{
  pdftitle={\mytitle},
  pdfauthor={Matthew Sun, Vinay Kothapally, Meng Yu, Chao Huang, Hao Zhang, Yixuan Zhang, Steve Yves}
}
\begin{document}
\bstctlcite{IEEEreferenceControl}
\maketitle

\AddToHookNext{shipout/foreground}{%
  \begin{tikzpicture}[remember picture,overlay]
    \node[anchor=south west,inner sep=0pt,align=left,
          text width=\textwidth,font=\footnotesize]
      at ([xshift=0.75in,yshift=0.25in]current page.south west)
      {\rule{5pc}{0.4pt}\\[-0.2ex]
       \textsuperscript{*}This work was performed while Matthew Sun was at Tencent Americas, Palo Alto, CA, USA. He is now at the School of Engineering and Applied Sciences, Columbia University, New York, NY, USA.\par
       \textsuperscript{\ensuremath{\dagger}}This work was performed while Hao Zhang was at Tencent Americas, Bellevue, WA, USA.};
  \end{tikzpicture}%
}

\begin{abstract}
Full-duplex dialogue systems, which listen while speaking, must distinguish a completed turn from a pause within a turn and an interruption that requests a turn from a brief acknowledgment or speech addressed to a third party. Yet existing conversational corpora provide limited control over these events and limited labels for their intent. We present a pipeline for synthesizing intent-labeled, two-channel conversational speech from relational event lists. A large language model (LLM) authors each event's speaker, text, conversational act, and attachment to an earlier event without predicting absolute timestamps. Events are synthesized independently, aligned with their source text to locate speech boundaries, and placed on a shared clock, so turn-taking landmarks are measured from the rendered signal while silence durations are specified or sampled from turn-taking distributions. The pipeline covers 42 phenomena across eight families in English and Mandarin, derives frame-level system actions from authored intent, and promotes diversity using small, diverse sets of prior examples and batch prompts that request alternatives with self-reported probabilities. Generation ablations show gains in the diversity dimensions targeted by each mechanism. On a four-action label space for taking, holding, releasing, and not holding the conversational floor (the right to continue a turn), a semantic voice-activity detector using only current and past audio reaches start-speaking and start-listening F1 scores of 0.819 and 0.802, respectively. When generating its own responses, the full-duplex speech model Moshi takes 0.85 of the reference turns after fine-tuning on the generated corpus, compared with 0.44 before fine-tuning. Its frame-level precision for predicting system-floor occupancy rises from 0.46 to 0.88. With reference conversational context supplied at each step, its frame-level floor F1 rises from 0.893 to 0.962. These results show that controlled synthesis can provide learnable and transferable supervision for full-duplex turn management.
\end{abstract}
\section{Introduction}
\label{sec:introduction}

\begin{figure*}[t!]
    \centering
    \includegraphics[width=\textwidth]{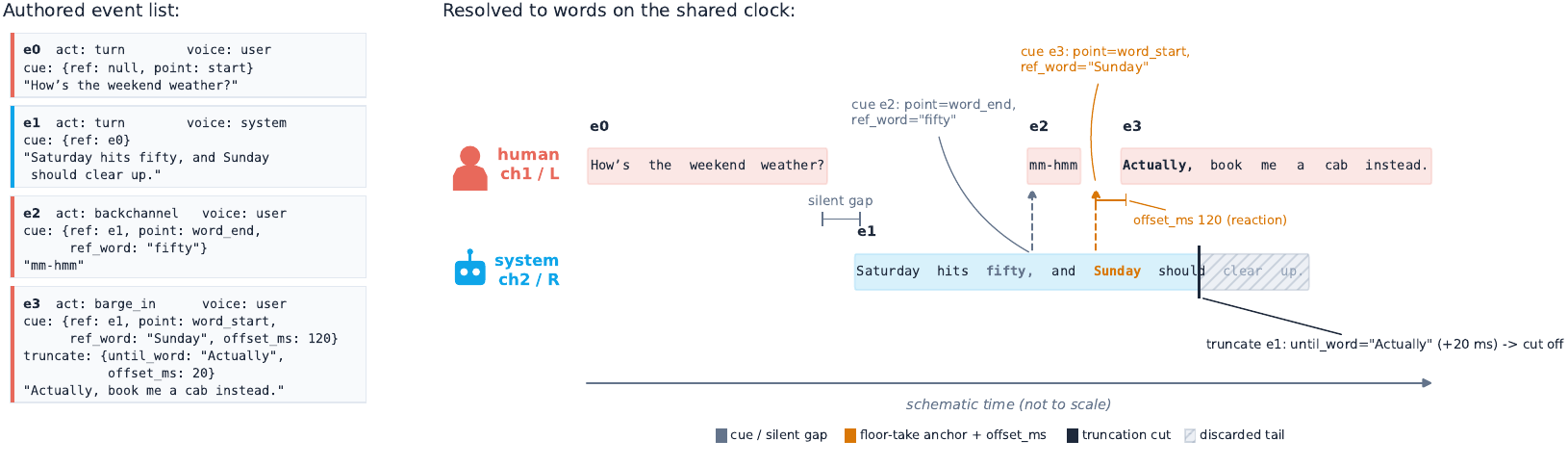}
    \caption{An example of how authored relational events become a resolved two-channel timeline. The event list specifies speaker, act, text, and cues relative to earlier events without absolute timestamps. Measured word boundaries place the user's backchannel after ``fifty'' and anchor a barge-in to ``Sunday''; the authored reaction offset and truncation rule determine the overlap and remove the interrupted tail of the system response. Channel~1 (ch1/L) carries human speech, and channel~2 (ch2/R) carries system speech; L and R denote the left and right audio channels.}
    \label{fig:event-schema}
\end{figure*}

Natural conversation does not alternate between complete utterances. Speakers pause without yielding, overlap, interrupt, offer brief acknowledgments (backchannels), repair their speech, and address people outside the main conversation. Across languages, responses often begin within a few hundred milliseconds of the preceding turn's end \cite{stivers2009universals}. Because planning a response can take longer than this interval, listeners must predict turn completion rather than react after silence begins \cite{levinson2015timing}. Turn-taking is a continuous decision based on prosody, syntax, semantics, and intent.

This decision is central to full-duplex systems, which listen while speaking. At each moment, the system must speak or continue listening. Silence may occur within a turn, while the user retains the floor (the right to continue the turn), or follow a completed turn that invites a response. Speech during a system response may be a floor-seeking interruption (barge-in) that should stop the response, a backchannel that should not, or speech addressed to a third party (side-talk). These cases can have similar durations and overlap but require different actions. Voice activity or a fixed silence threshold cannot distinguish them.

Spoken dialogue models learn these behaviors from synchronized two-speaker audio. The dialogue generative spoken language model (dGSLM) learns overlap and backchannels from two-channel Fisher \cite{cieri2004fisher} conversations \cite{nguyen2023dgslm}, while Moshi combines recorded conversations with synthetic dialogue \cite{defossez2024moshi}. These sources leave a supervision gap. Real corpora contain natural timing but do not identify the intent of each overlap or balance phenomena by construction. Audio mixtures for speaker diarization---identifying who speaks when---reproduce overlap statistics but assign the same activity pattern to barge-ins, backchannels, and other overlapping speech \cite{landini2022simulated}. Text dialogue synthesis controls content but not its acoustic timeline.

We address this gap with a pipeline for controlled, intent-labeled, two-channel speech in English and Mandarin. A large language model (LLM) represents each conversation as relational events that specify the speaker, text, conversational act (the event's role in the exchange), and attachment to an earlier event, without absolute timestamps. Each event is synthesized and then forced-aligned---matched to its source text to measure word or character boundaries---and placed on a shared clock using word-anchored cues and inserted silence. Speech landmarks are measured from the rendered signal; silence durations are specified or sampled.

\Cref{fig:event-schema} shows this process. Authored acts, speakers, and cues combine with aligned word boundaries to place a backchannel and barge-in on the two-channel clock and truncate the interrupted system turn.

The representation preserves the intent behind ambiguous acoustic events. It supports frame-level system actions---take the floor, speak, release the floor, or listen---and alternative label schemes without resynthesizing audio. The timing model samples turn gaps, within-speaker pauses, and interruption reactions from overlapping distributions grounded in conversation research, preventing duration from serving as the sole cue. The corpus covers 42 phenomena across eight families in English and Mandarin.

We contribute a pipeline with the following key features: (1) a relational event representation for controlled two-channel dialogue; (2) a rendering method that aligns labels by measuring speech landmarks and specifying silences; (3) tools for improving lexical diversity and conversational chronemics (the timing of pauses, overlaps, and turn transitions); and (4) intent-derived labels that distinguish barge-ins from backchannels.

\section{Preliminaries}
\label{sec:preliminaries}

This section defines the turn-taking concepts and representation used in the paper.

\subsection{Floor, Turns, and Overlap}

The \emph{floor} is the recognized right to continue a conversational turn. Conversation analysis describes turn-taking as a coordinated system that limits gaps and overlaps rather than eliminating them \cite{sacks1974systematics}. A \emph{floor transfer} moves this right between speakers. The floor-transfer offset (FTO) is the signed difference between the end of one speaker's activity and the start of the next \cite{heldner2010pauses}. Positive FTO denotes a gap; negative FTO denotes overlap.

A floor-transfer gap differs from a \emph{within-speaker pause}, in which a silent speaker retains the floor and continues the same turn. Their duration distributions overlap, so elapsed silence does not determine whether the floor was yielded. This ambiguity motivates continuous models that predict future joint voice activity instead of waiting for a detected endpoint \cite{ekstedt2022vap}.

Overlap depends on intent. A \emph{barge-in} contests another speaker's floor; it succeeds when the floor-holder yields and fails when the interrupter backs off. A \emph{backchannel}, such as ``mm-hmm,'' acknowledges a speaker without requesting the floor. \emph{Side-talk} addresses a third party and should not trigger a system response. These acts can produce similar activity patterns, so their labels follow authored intent rather than overlap geometry.

\subsection{Relational Event Representation}

A conversation is an ordered set of speech \emph{events}. Each event records its speaker, channel, text, act, and relational timing cue. Channel~1 contains the user and any side-talker; channel~2 contains the system. The core acts are:

\begin{itemize}
    \item \texttt{turn}: begins a new floor-holding turn;
    \item \texttt{continue}: resumes the same speaker after a within-turn pause;
    \item \texttt{backchannel}: acknowledges without taking the floor;
    \item \texttt{barge\_in}: contests the floor and truncates the referenced event when \texttt{interrupts} is \texttt{true}; and
    \item \texttt{aside}: addresses a third party without transferring the floor.
\end{itemize}

An event cue references an earlier event and anchors to its start, effective end (the end after any truncation), or the measured boundary of a named word. It can add a gap and signed offset. Backward-only references form a directed acyclic graph resolved in event order. An author can therefore place an interruption after a word without predicting when text-to-speech (TTS) synthesis will produce it.

\subsection{From Events to Training Labels}

The conversation is generated as one JavaScript Object Notation (JSON) object, but each event is rendered and forced-aligned separately. Word-relative starts and interruption cuts use the measured boundaries. Turn gaps and within-speaker pauses are silence durations sampled when the author leaves them unspecified. The clips are placed at sample-accurate offsets on a shared two-channel clock.

Labels are defined from the system's perspective on an 80-ms grid, which can be adjusted for another model architecture. The four floor actions are \emph{take floor} (the start-speaking label), \emph{speaking} (holding the floor), \emph{release floor} (start-listening), and \emph{listening} (not holding the floor). Start-speaking and start-listening mark single-frame transitions; speaking and listening label spans. In finer label spaces, ambient silence may form a separate state. Authored intent resolves ambiguous regions: a successful user barge-in triggers release, while a backchannel, side-talk, or failed barge-in does not. Because labels derive from the event graph, word times, and timeline, other frame rates or act taxonomies require no resynthesis.
\section{Approach}
\label{sec:approach}

Our pipeline separates conversation \emph{authoring} from acoustic \emph{realization}. Authoring determines what happens and the intent behind it: the speakers, text, conversational acts, and relations among events. Realization determines when each sound occurs by rendering the events with TTS, measuring their word timings with a forced aligner, and resolving them onto a shared clock. This separation avoids asking either the authoring model or the speech synthesizer to predict absolute timestamps.

\begin{figure*}[t]
    \centering
    \includegraphics[width=\textwidth]{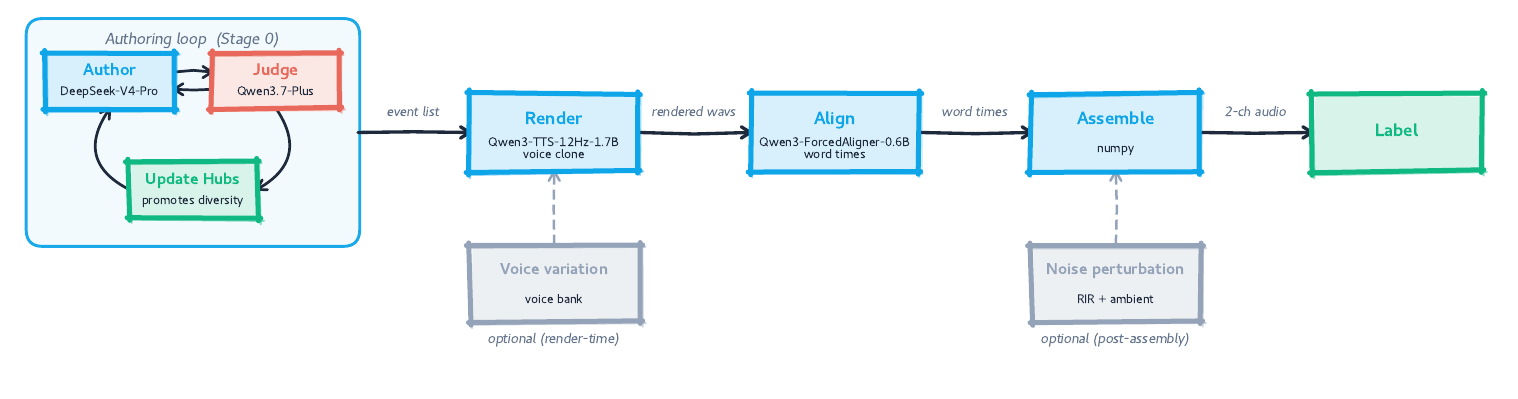}
    \caption{End-to-end generation pipeline. The authoring loop creates and validates relational event lists while updating the diversity hubs. Accepted events are rendered independently, forced-aligned, assembled on a shared two-channel clock, and converted to frame-level labels. Voice variation and noise perturbation are optional and do not alter the authored event structure. RIR denotes a room impulse response, which models how a room modifies sound; wavs denotes waveform audio files, and 2-ch denotes two-channel audio.}
    \label{fig:pipeline-blocks}
\end{figure*}

\Cref{fig:pipeline-blocks} summarizes the complete path from authored events to labeled two-channel audio, including the optional acoustic-variation stages.

The prompt templates used for conversation authoring are reproduced in \hyperref[sec:appendix]{Appendix~\ref*{sec:appendix}}.

\subsection{Scenario and Conversation Specification}

The scenario space contains 42 phenomena in eight families: ordinary turn-taking, overlap and interruption, backchannels, disfluency and repair, addressee and multi-party behavior, task structure, affect, and edge cases. Each phenomenon has a language-independent \texttt{scenarioID}, a short description of its floor mechanics (the pattern of floor holding, transfer, and overlap), and rules that generated conversations must preserve. Language-specific configuration overlays for English and Mandarin localize the authoring instructions and speaker descriptions without changing the timing policy.

\begin{figure}[t]
    \centering
    \includegraphics[width=\columnwidth]{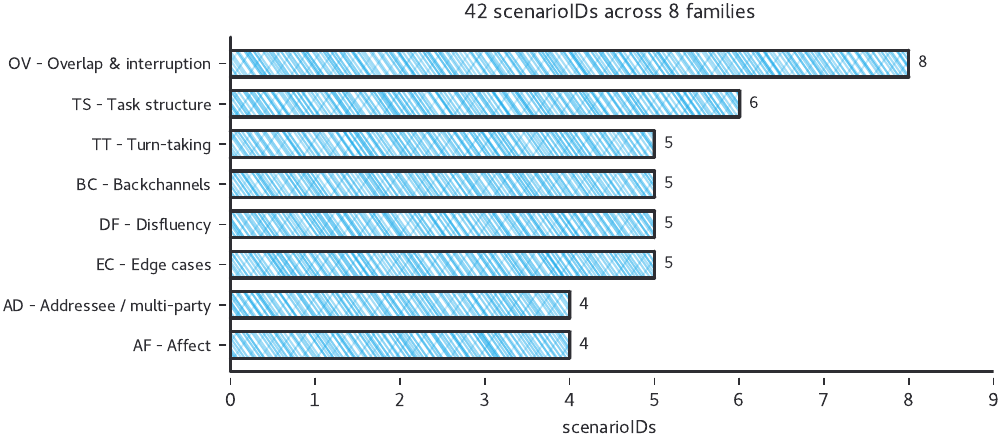}
    \caption{Coverage of the scenario registry. The 42 scenario identifiers span eight families, with the largest allocation devoted to overlap and interruption phenomena.}
    \label{fig:scenario-family-coverage}
\end{figure}

\Cref{fig:scenario-family-coverage} shows the distribution of the 42 scenario identifiers across these eight families.

For each scenario, the authoring model, DeepSeek-V4-Pro \cite{deepseekai2026v4}, generates one complete conversation JSON object. The object contains a topic description and an ordered event list using the representation in Section~\ref{sec:preliminaries}. Timing remains relational: an event can follow the effective end of a referenced event, begin at a named word boundary, or overlap it by a signed offset. The model therefore expresses an interruption as, for example, ``start shortly after the first occurrence of this word,'' rather than estimating seconds or audio frames.

Each draft passes staged quality checks before synthesis. Static validation checks unique identifiers, legal acts and cue fields, backward-only references, and the presence and occurrence count of every anchor or truncation word. A timeline precheck uses estimated word durations and the same dependency resolver as final assembly to reject accidental same-channel overlaps. Text filters detect repeated, excessively long, symbol-heavy, or otherwise unsuitable utterances. A deterministic normalizer rewrites digits, clock times, and symbols into spoken forms so that written and aligned token counts remain consistent. Optionally, a second language model, Qwen3.7-Plus \cite{qwen2026max37plus}, acts as a judge to check semantic coherence and whether a valid anchor is also a sensible landmark for the intended event; failed drafts are regenerated with the judge's feedback.

\subsection{Render, Align, and Assemble}

Every accepted event is synthesized independently with Qwen3-TTS \cite{hu2026qwen3tts}, conditioned on a reference recording for its assigned speaker. Independent synthesis is important because the TTS system is free to choose natural prosody and duration; the pipeline measures the resulting timing instead of forcing the waveform to match a hand-authored schedule. Separate reference voices are used for the user, system, and any side-talker.

Next, each event waveform is forced-aligned with its source text using Qwen3-ForcedAligner \cite{shi2026qwen3asr}. The aligner returns clip-relative start and end times for every word. Mandarin is matched at the character level, with consecutive characters grouped when an anchor contains more than one character. Very short clips, such as one-word backchannels, receive temporary trailing padding for alignment stability; this padding is removed before assembly. Each waveform is then trimmed to its measured spoken span, and its word times are rebased so that the first spoken unit begins at zero.

Assembly resolves each event in dependency order. For event $i$, its absolute start time is

\[
t_i = t_{r(i)} + p_{r(i)} + g_i + o_i,
\]

where $r(i)$ is the referenced event, $p_{r(i)}$ is the measured cue point within that event, $g_i$ is the inserted gap, and $o_i$ is a small signed offset from the landmark. The cue point can be the referenced event's start, its effective end after truncation, or a measured word start or end. A positive gap inserts silence, while a negative gap creates intentional overlap. The same rule represents normal replies, held-floor pauses, low-latency turn transitions, backchannels, and interruptions.

For a successful \texttt{barge\_in}, assembly also shortens the interrupted event. The cut time is derived from a word boundary in the interrupting utterance plus its authored offset and then clamped to the target event's time span. The waveform is briefly faded at the cut to avoid a discontinuity. A failed barge-in retains the overlap but does not truncate the floor-holder. Events are finally copied at sample-accurate positions into two zero-filled buffers: channel~1 for the user and side-talkers, and channel~2 for the system. Same-channel speech is serialized, while cross-channel overlap is preserved by design.

Locations inside speech, including word-relative starts and interruption cuts, come from forced alignment. Locations between speech segments are inserted silences. Consequently, a change in speaking rate or voice does not invalidate the timeline or its labels.

\subsection{Natural Timing and Intent-Derived Labels}

When the author supplies a gap or reaction time, assembly preserves it. Otherwise, the pipeline samples a value according to the event type. Turn-yield gaps are sampled from a reconstruction of observed floor-transfer offsets; within-speaker pauses use a separate pause distribution; and barge-in offsets use reaction-time ranges associated with correction, redirection, impatience, cooperative completion, or failed interruption. These are fallback distributions rather than global overrides.

The pause and turn-transfer distributions deliberately overlap. A long within-turn pause can exceed a short reply gap, matching the ambiguity found in real conversation and preventing a model from solving the task with a single duration threshold. Negative transfer offsets also produce natural overlap around turn boundaries. The sampled value is recorded in the resolved event data so the resulting distribution can be audited against literature-derived and real-speech references.

After the timeline is resolved, a converter derives the system-side labels described in Section~\ref{sec:preliminaries}. Acoustic activity determines speaking spans and boundary frames, while authored acts determine who owns or contests the floor. A successful user interruption makes the system release the floor. A backchannel, aside, background speech, or failed barge-in leaves the system's floor state unchanged. This intent-aware pass distinguishes regions that would be identical under voice-activity labels alone. Because the converter operates on stored events, word alignments, and timestamps, a different frame interval or act taxonomy requires only a new converter, not new TTS output.

We present illustrations of two conversations under three possible label granularities in \hyperref[sec:label-examples]{Appendix~\ref*{sec:label-examples}}.

\begin{figure*}[t]
    \centering
    \includegraphics[width=\textwidth]{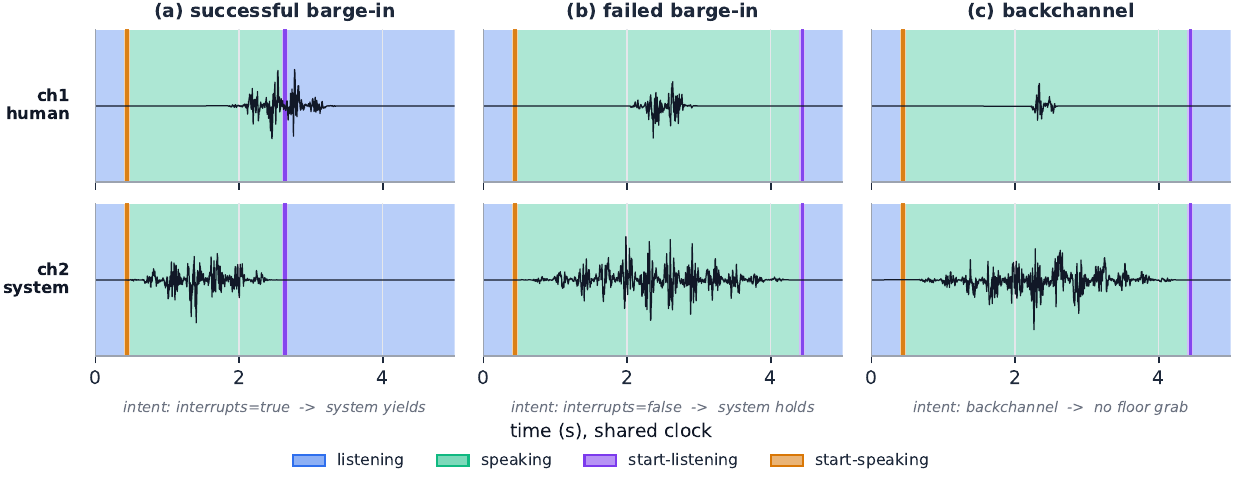}
    \caption{Illustrating intent-aware labels for three acoustically similar overlap patterns. A successful barge-in transfers the floor and causes the system to stop, whereas a failed barge-in and a backchannel leave the system's floor ownership unchanged. Colored regions show system-side frame labels on the shared two-channel clock.}
    \label{fig:intent-triptych}
\end{figure*}

\Cref{fig:intent-triptych} illustrates why activity alone is insufficient: similar cross-channel overlaps require different system actions when they represent a successful barge-in, failed barge-in, or backchannel.

\subsection{Few-Shot Diversity Hubs}

Generating conversations independently can lead to mode collapse, or repetitive output: later samples reuse familiar topics, act sequences, and timing patterns because the authoring model cannot see what the run has already produced. The hub design is inspired by ConvoGen's iterative sampling, which updates few-shot context---the small set of examples included in the prompt---with experiences generated during earlier calls \cite{gody2025convogen}. Our use differs: hub members serve as negative memory that the authoring model is instructed not to repeat, rather than examples to imitate. We maintain two such few-shot \emph{hubs}. Each hub stores a coreset, a small subset of prior conversations selected to cover diverse outputs, for every \texttt{scenarioID}. The structural hub tracks floor mechanics, while the semantic hub tracks topic and meaning. Both hubs select examples with the same online cover algorithm, which incrementally maintains a diverse subset as candidates arrive, but they use different conversation representations and distance functions. Topic seeding is complementary: it draws subjects without replacement from a concept bank derived from Wikipedia's Level~5 Vital Articles list \cite{wikipedia2026vital5}, whereas the hubs react to the content and structure the model generates.

\subsubsection{Structural Hub: Floor Fingerprints}

The structural hub represents a conversation by how its speakers manage the floor rather than by its words. Its continuous representation is a 27-dimensional feature vector containing per-act counts, the histogram of barge-in types, the number and rate of floor transfers, summary statistics of gaps and overlaps, the fraction of the conversation containing overlap, barge-in reaction-time statistics, and the number of truncations. The same feature extraction routine is used by the corpus analysis tools, so the geometry used to steer generation agrees with the structural diversity reported during evaluation.

Because these features use different units and scales, each column is standardized using the running mean and standard deviation of all conversations observed for that scenario. Hub members retain their raw vectors and are re-standardized whenever the running statistics change. Euclidean feature distance is then compressed to the interval $[0,1]$ relative to the mean pairwise feature distance over the current members together with the candidate. This shared normalization applies to both member--member and candidate--member distances, preventing a single extreme timing feature from dominating the comparison.

The hub also preserves event order through an act-sequence representation. Each event becomes an act token, with a barge-in's subtype included so that, for example, a correction and a redirect remain distinct. Sequence distance can use one minus act-level ROUGE-L \cite{lin2004rouge}, a similarity score based on the longest common subsequence, or a global alignment that additionally penalizes skipped turns. We use the \texttt{blend} structural distance with weight $w=0.75$ on the ordered act-sequence distance and weight $1-w=0.25$ on the normalized feature distance. Thus, two conversations are close only when both their aggregate timing regimes and their progression of floor actions are similar. In the prompt, structural members are shown as compact floor fingerprints---act sequence, interruption opener and type, reaction offset, anchor word, and truncation---rather than as an opaque vector of numbers.

For example, the events in \Cref{fig:event-schema} would be represented for the structural hub as \texttt{turn} $\rightarrow$ \texttt{turn} $\rightarrow$ \texttt{backchannel} ``mm-hmm'' on ``fifty'' $\rightarrow$ \texttt{barge\_in/redirect} opener ``Actually,'' react 120\,ms on ``Sunday,'' truncate ``Actually'' +20\,ms. This compact form drops the topic and full utterances while retaining the act order and the timing and interruption choices that define the conversation's rhythm.

\begingroup
\makeatletter
\renewcommand{\subsubsection}{\@startsection{subsubsection}{3}{\z@}%
  {-6pt}{-1em}{\normalsize\bf}}
\makeatother
\setlength{\parskip}{0pt}
\subsubsection{Semantic Hub: Meaning Embeddings}

The semantic hub targets a different failure mode: conversations can use different floor patterns while repeatedly discussing the same ideas, or can paraphrase one another with low surface word overlap. For each conversation, all event text is concatenated in order and encoded by BGE-M3 \cite{chen2024bgem3} as a 1024-dimensional sentence embedding, a numerical vector representing its meaning. BGE-M3 provides a shared multilingual representation for English and Mandarin. The vectors are normalized to unit Euclidean (L2) length, so Euclidean distance is monotonic with cosine distance and no additional feature standardization is required. Paraphrases remain close even when their exact words differ, while conversations about different content are farther apart.

Semantic embeddings are computed outside the authoring process because the embedding model and generation stack have incompatible runtime dependencies. Existing conversations are loaded from a cache keyed by sample identifier and content hash. New conversations are sent to a persistent embedding service so that the semantic cover can be updated during the same generation run. This path fails open: if an embedding is absent or the service is unavailable, generation continues using the structural hub rather than discarding the sample. Semantic hub members are presented to the authoring model as short content exemplars, which exposes their meaning without exposing embedding coordinates.

\subsubsection{Online Coreset Maintenance}

Both hubs use the streaming max-min cover in Algorithm~1: the retained set favors candidates that are far from their nearest retained neighbor. Here, $C$ is the retained coreset, $n$ is the number of candidates observed, and $M$ is its current capacity. The distance $d$ is the hub-specific structural or semantic distance defined above; for the structural hub it uses the current running standardization statistics. Existing corpus records are replayed through the same update on startup. Because the cover capacity is bounded between 5 and 32, an update costs $O(M^2)$ and is effectively constant in total corpus size.
\endgroup

\begin{figure}[t]
\centering
\begin{minipage}{0.98\columnwidth}
\small
\hrule
\vspace{3pt}
\textbf{Algorithm 1: Streaming max-min coreset update}
\begin{algorithmic}[1]
\Require candidate $x$, cover $C$, observed count $n$
\Require bounds $M_{\min}=5$, $M_{\max}=32$, distance $d$
\State $n \gets n+1$; update running statistics with $x$
\State $M \gets \min(M_{\max},\max(M_{\min},\lceil\sqrt{n}\rceil))$
\If{$|C| < M$}
    \State \Return $C\cup\{x\}$
\EndIf
\State $s_x \gets \min_{c\in C} d(x,c)$ \Comment{new isolation}
\ForAll{$c\in C$}
    \State $s_c \gets \min_{c'\in C\setminus\{c\}} d(c,c')$
\EndFor
\State $c^\star \gets \arg\min_{c\in C}s_c$
\If{$s_x > s_{c^\star}$} \Comment{above crowding floor}
    \State $C \gets (C\setminus\{c^\star\})\cup\{x\}$
\EndIf
\State \Return $C$
\end{algorithmic}
\vspace{3pt}
\hrule
\end{minipage}
\end{figure}

Hub membership affects only future prompts. Every conversation that passes the quality checks is written to the corpus whether or not it enters either cover. This distinction avoids rejection loops and the saturation problem of a fixed similarity threshold: as a scenario fills, the hubs continue to provide diverse few-shot context without making valid output progressively harder to accept.

\subsection{Probability-Verbalized Batch Authoring}

The hubs increase diversity across successive authoring calls, but a single call can still return the model's most typical output. We therefore also adapt Verbalized Sampling, a training-free prompting method that asks a language model to state a distribution over several responses rather than produce only one response \cite{zhang2025verbalized}. In our setting, the prompt requests a batch of five complete conversations and a self-reported probability for each conversation. The probabilities make the model expose lower-probability alternatives that ordinary single-response decoding tends to suppress.

All members of a batch receive the same scenario constraints and, when enabled, the same seeded topic. The prompt explicitly asks them to explore distinct interpretations, subtopics, wording, and floor-management patterns while remaining valid instances of that scenario. Generating the alternatives together lets the model contrast them within its context. For a broad topic, for example, different batch members can focus on different applications or subthemes instead of independently converging on the topic's most familiar framing. The probability values serve as a prompting device and are not used as corpus weights or acceptance scores.

Each returned conversation is then handled as an ordinary candidate: it is normalized, statically validated, checked for timeline conflicts and content failures, and written only if it passes the same quality gates. Accepted batch members update the structural and semantic hubs one at a time, so batch authoring and hub steering compose cleanly. Probability verbalization promotes within-call variation, while the hubs provide memory across calls and prevent later batches from revisiting the same structural and semantic regions. The technique is optional, allowing its contribution to lexical, semantic, and floor-mechanics diversity to be isolated experimentally.

\subsection{Acoustic Variation and Reusable Outputs}

Acoustic variation is applied independently of authoring. At render time, speaker reference recordings can be sampled without replacement from a voice bank grouped by language, sex, and age. Its public-source recordings come from LibriSpeech, VCTK, the clean-speech portion of the Interspeech 2020 Deep Noise Suppression Challenge, and DiDiSpeech, supplemented by an internal Mandarin TTS corpus \cite{panayotov2015librispeech,yamagishi2019vctk,reddy2020dns,guo2020didispeech}. After assembly, time-varying recorded noise can be mixed into the user channel while retaining a clean parallel copy and leaving the system channel unchanged. The perturbations use the Kaldi Room Impulse Response and Noise collection (OpenSLR-28), which contains real and simulated room responses as well as spatially diffuse noise and noise from localized sources; the latter include material from the MUSAN music, speech, and noise corpus \cite{openslr28,ko2017augmentation,snyder2015musan}. Both operations preserve event structure and labels, allowing the same authored corpus to support multiple acoustic conditions.

Each completed sample stores the authored event object, the two-channel waveform, the transcript of speech that actually played after truncation, event-level timestamps, and an optional resolved timeline with word boundaries. These artifacts make every sample auditable and allow the corpus to be re-rendered, relabeled, or analyzed without repeating the full authoring process.
\section{Evaluation}
\label{sec:evaluation}

We evaluate the pipeline at three levels. First, controlled generation ablations test whether the prompt-side diversity mechanisms change the produced corpus. Second, an internal semantic voice-activity detector (semantic VAD), which predicts intended floor actions rather than only acoustic speech activity, tests whether the authored labels are recoverable from audio and remain stable under acoustic perturbation. Finally, fine-tuning Moshi tests whether the corpus improves an independently designed full-duplex generative model.

\subsection{Generation-Pipeline Ablations}

The following experiments are targeted tests rather than a large hyperparameter sweep. Each holds the scenario, sample count, and remaining generation settings fixed while changing one intervention. Each experimental condition (arm) contains 45 English conversations. Expectation-Adjusted Distinct (EAD-$n$) measures the variety of $n$-token sequences while adjusting for sample length \cite{liu2022distinct}; $n=1,2,3$ denotes unigrams, bigrams, and trigrams. Lexical tokens are words or characters, whereas act-sequence tokens represent conversational acts. The unique-types metric counts distinct lexical tokens. The Vendi score measures the effective number of distinct samples from a pairwise similarity matrix \cite{friedman2022vendi}. We compute structural Vendi in the standardized feature space and semantic Vendi with a radial basis function (RBF, or Gaussian) kernel over meaning embeddings. Nearest-neighbor (NN) mean distance measures the average distance to each conversation's closest neighbor in the standardized structural space. Higher EAD, Vendi, and NN distance indicate more variety; lower mean pairwise ROUGE-L \cite{lin2004rouge} indicates less repetition. Tables abbreviate act sequence as Act-seq.; $\Delta$ denotes the change from the baseline, expressed as a relative percentage in the ablation tables.

\paragraph{Structural hub.}
We compare the default \texttt{blend} structural hub against a no-hub baseline on AF-02 (amused speech, including laughter while speaking), with no topic seeding or semantic hub. \Cref{tab:structural-ablation} reports only structural measurements from this comparison. Without the hub, generation drops barge-ins and truncations entirely and collapses onto a turn-plus-backchannel pattern. Enabling \texttt{blend} restores the interruption repertoire and clearly increases structural diversity: feature-space Vendi rises by 64.2\%, nearest-neighbor distance by 95.0\%, and act-sequence EAD-2 by 185.7\%, while pairwise act-sequence ROUGE-L decreases. As shown in \Cref{tab:structural-ablation}, the gain therefore reflects broader floor-mechanics coverage rather than only longer conversations or more events.
\begin{table}[t]
\caption{Structural-hub ablation on AF-02. \texttt{Blend} combines feature and act-sequence distances. Arrows indicate the direction of better performance.}
\label{tab:structural-ablation}
\centering
\small
\resizebox{\columnwidth}{!}{%
\begin{tabular}{lrrr}
\hline
\textbf{Structural metric} & \textbf{No hub} & \textbf{Blend} & \textbf{$\Delta$} \\
\hline
Barge-ins (total) & 0 & 46 & -- \\
Truncations (total) & 0 & 42 & -- \\
Backchannels (total) & 48 & 86 & +79.2\% \\
Feature-space Vendi $\uparrow$ & 4.29 & 7.05 & +64.2\% \\
NN mean distance $\uparrow$ & 1.37 & 2.66 & +95.0\% \\
Act-seq. ROUGE-L $\downarrow$ & 0.861 & 0.769 & -10.7\% \\
Act-seq. EAD-2 $\uparrow$ & 0.350 & 1.000 & +185.7\% \\
Events / conversation & 8.93 & 14.11 & +58.0\% \\
Floor transfers / conversation & 6.80 & 11.76 & +72.9\% \\
\hline
\end{tabular}}
\end{table}

\paragraph{Semantic hub.}
We isolate the semantic hub on AF-02 with identical topic seeding and the structural hub disabled. \Cref{tab:semantic-ablation} therefore reports only semantic and lexical measurements. Semantic Vendi increases by 3.6\%, showing additional meaning dispersion beyond the already strong distinct-topic baseline. Unique types and all length-adjusted EAD metrics increase, while pairwise ROUGE-L increases slightly. The surface-lexical evidence is therefore mixed, while the target semantic axis shows a modest positive gain.

\begin{table}[t]
\caption{Semantic-hub ablation on AF-02 with identical topic seeding in both arms.}
\label{tab:semantic-ablation}
\centering
\small
\resizebox{\columnwidth}{!}{%
\begin{tabular}{llrrr}
\hline
\textbf{Axis} & \textbf{Metric} & \textbf{Seeded} & \textbf{+Hub} & \textbf{$\Delta$} \\
\hline
Semantic & RBF-Vendi $\uparrow$ & 9.402 & 9.738 & +3.6\% \\
Lexical & Unique types $\uparrow$ & 1449 & 1732 & +19.5\% \\
Lexical & Pairwise ROUGE-L $\downarrow$ & 0.141 & 0.150 & +6.0\% \\
Lexical & EAD-1 $\uparrow$ & 0.681 & 0.780 & +14.4\% \\
Lexical & EAD-2 $\uparrow$ & 1.060 & 1.079 & +1.7\% \\
Lexical & EAD-3 $\uparrow$ & 1.208 & 1.241 & +2.7\% \\
\hline
\end{tabular}}
\end{table}

\paragraph{Probability verbalization.}
We isolate batch authoring on TT-01 (ordinary back-and-forth turn-taking) with one fixed topic and both hubs disabled. The verbalized arm requests five candidates with self-reported probabilities in each call \cite{zhang2025verbalized}. \Cref{tab:prob-ablation} reports both content and structural diversity. Lexical EAD-2 rises by 37.6\%, pairwise lexical ROUGE-L falls by 37.8\%, and semantic Vendi rises by 68.7\%. Structural Vendi also increases, but nearest-neighbor distance decreases, act-sequence measurements remain flat, and gap spread narrows. Probability verbalization therefore strongly diversifies wording and meaning, while its effect on floor mechanics is mixed.

\begin{table*}[t]
\caption{Probability-verbalization ablation on TT-01 with one fixed topic and no diversity hubs.}
\label{tab:prob-ablation}
\centering
\small
\begin{tabular}{llrrr}
\hline
\textbf{Axis} & \textbf{Metric} & \textbf{Single} & \textbf{Batch} & \textbf{$\Delta$} \\
\hline
Lexical & Unique types $\uparrow$ & 631 & 874 & +38.5\% \\
Lexical & Pairwise ROUGE-L $\downarrow$ & 0.280 & 0.174 & -37.8\% \\
Lexical & EAD-2 $\uparrow$ & 0.724 & 0.996 & +37.6\% \\
Semantic & RBF-Vendi $\uparrow$ & 2.262 & 3.815 & +68.7\% \\
Structural & Feature-space Vendi $\uparrow$ & 3.442 & 3.776 & +9.7\% \\
Structural & NN mean distance $\uparrow$ & 0.704 & 0.607 & -13.8\% \\
Structural & Act-seq. ROUGE-L $\downarrow$ & 0.935 & 0.946 & +1.2\% \\
Structural & Act-seq. EAD-2 $\uparrow$ & 1.000 & 1.000 & 0.0\% \\
Structural & Gap standard deviation (s) $\uparrow$ & 0.509 & 0.353 & -30.7\% \\
\hline
\end{tabular}
\end{table*}

Together, the experiments support complementary mechanisms: the default structural hub prevents floor-pattern collapse, the semantic hub adds meaning spread beyond topic seeding, and probability verbalization substantially increases within-call content diversity while requiring monitoring of its timing distribution.

\subsection{Internal Semantic-VAD Benchmark}

We evaluate four runs on the English-plus-Mandarin speech corpus: the 4-token \texttt{core} and 13-token \texttt{all\_merged} label spaces, each on clean and noisy audio. Here, a token is one categorical frame label, not a word or an act-sequence token. The internal model is entirely causal: each prediction uses only current and past audio. All runs use the same split of 19,575 training clips and 2,756 validation clips, with no speakers or conversations shared between the two sets. Validation comprises 910,254 frame labels, with one prediction every 80 ms. In the noisy condition, only the user channel receives the headset-microphone noise perturbation (recorded noise modified by room impulse responses); the system channel, labels, and split are identical to the clean condition.

\paragraph{Four-token core space.}
The core labels are start-speaking, speaking, start-listening, and listening. We report classification accuracy and per-class F1, the harmonic mean of precision and recall. In this and subsequent classification tables, $\Delta$ denotes an absolute score change. \Cref{tab:semantic-vad-core} gives the results. Clean accuracy is 0.9932, with speaking and listening near 0.995 F1 and the two sparse boundary classes at 0.8186 and 0.8019. User-channel noise changes accuracy by only -0.0003; start-speaking F1 changes by -0.0023 and start-listening F1 by -0.0061. Start-speaking latency measures onset timing relative to the reference. Its mean is 6.2 ms on clean audio and 5.2 ms on noisy audio, while the median and 90th percentile are zero in both cases.

\begin{table}[t]
\caption{Exact held-out results for the 4-token core semantic-VAD runs.}
\label{tab:semantic-vad-core}
\centering
\small
\begin{tabular}{lrrr}
\hline
\textbf{Core metric} & \textbf{Clean} & \textbf{Noisy} & \textbf{$\Delta$} \\
\hline
Accuracy & 0.9932 & 0.9929 & -0.0003 \\
Speaking F1 & 0.9953 & 0.9951 & -0.0002 \\
Listening F1 & 0.9955 & 0.9953 & -0.0002 \\
Start-speaking F1 & 0.8186 & 0.8163 & -0.0023 \\
Start-listening F1 & 0.8019 & 0.7958 & -0.0061 \\
\hline
\end{tabular}
\end{table}

\paragraph{Thirteen-token merged space.}
The merged space retains finer turn-taking acts but combines user-originated and companion-originated interruptions under \emph{yield}, and their backchannels under \emph{hold-backchannel}. A companion is a third person on the human channel. These distinctions concern who acts, not whom the speech addresses. The additional labels distinguish system-floor states (\emph{contested-floor}: non-backchannel speech overlaps the system; \emph{hold-backchannel}: a human backchannel overlaps it; \emph{continue}: a silent pause while the system retains the floor), listening states (\emph{system-backchannel}: the system acknowledges without taking the floor; \emph{continue-listening}: a human pauses without yielding; \emph{reply-gap}: silence during which a system reply is expected; \emph{silence}: ambient silence with no reply expected), and transitions (\emph{yield}: the system releases the floor after an interruption; \emph{barge-in}: the system takes the floor while a human is speaking). \Cref{tab:semantic-vad-merged} reports every class from the clean and noisy \texttt{all\_merged} records. Clean accuracy is 0.9613. Speaking and listening remain above 0.98 F1; contested-floor and hold-backchannel exceed 0.90; and the rare barge-in class is lowest at 0.4325 F1 on 209 frames.

\begin{table}[t]
\caption{Exact held-out results for the 13-token merged semantic-VAD runs.}
\label{tab:semantic-vad-merged}
\centering
\small
\resizebox{\columnwidth}{!}{%
\begin{tabular}{lrrr}
\hline
\textbf{Merged metric} & \textbf{Clean} & \textbf{Noisy} & \textbf{$\Delta$} \\
\hline
Accuracy & 0.9613 & 0.9525 & -0.0088 \\
Speaking F1 & 0.9840 & 0.9822 & -0.0018 \\
Listening F1 & 0.9860 & 0.9816 & -0.0044 \\
Contested-floor F1 & 0.9177 & 0.8901 & -0.0276 \\
Hold-backchannel F1 & 0.9061 & 0.8568 & -0.0493 \\
System-backchannel F1 & 0.8888 & 0.8766 & -0.0123 \\
Continue F1 & 0.8659 & 0.8626 & -0.0032 \\
Silence F1 & 0.8277 & 0.7928 & -0.0348 \\
Continue-listening F1 & 0.8123 & 0.7575 & -0.0548 \\
Start-speaking F1 & 0.8104 & 0.8084 & -0.0021 \\
Reply-gap F1 & 0.8017 & 0.7045 & -0.0971 \\
Yield F1 & 0.7853 & 0.7847 & -0.0006 \\
Start-listening F1 & 0.7767 & 0.7680 & -0.0087 \\
Barge-in F1 & 0.4325 & 0.4045 & -0.0281 \\
\hline
\end{tabular}}
\end{table}

Noise has its largest effect on labels that require detecting silence on the perturbed user channel: reply-gap falls by 0.0971 F1, continue-listening by 0.0548, and silence by 0.0348. The principal floor transitions remain stable: start-speaking F1 decreases by 0.0021, start-listening F1 by 0.0087, and yield F1 by 0.0006. This localizes the degradation to the expected silence-sensitive classes rather than the core floor decision.

Merging the user-originated and companion-originated variants of yield and hold-backchannel also improves classification performance. Relative to the corresponding 15-token clean run, merging raises accuracy from 0.9584 to 0.9613, hold-backchannel F1 from 0.7120 to 0.9061, and yield F1 from 0.7211 to 0.7853. On noisy audio, it raises accuracy from 0.9506 to 0.9525, hold-backchannel F1 from 0.7005 to 0.8568, and yield F1 from 0.7196 to 0.7847. The merged space is therefore the stronger act-level target in both acoustic conditions.

\subsection{Moshi Benchmark}

We fine-tune the Moshiko English checkpoint of Moshi \cite{defossez2024moshi} on the English corpus, with a reference voice assigned to each speaker for each conversation. The speaker- and conversation-disjoint split contains 5,906 training clips and 890 validation clips. We evaluate both free generation, where Moshi receives only the user channel and generates its own assistant stream, and teacher forcing, where the reference human and system streams provide the context for each one-step prediction. Moshi can produce speech in short bursts separated by brief inactive gaps. Before extracting start-speaking (SS) and start-listening (SL) landmarks, we therefore merge inactive gaps of at most 480 ms in both the predicted and reference activity streams, treating the surrounding speech as one continuous turn. We then score the landmarks at tolerances of zero to three frames (0--240 ms on the 80-ms grid). Turn coverage is the fraction of reference turns taken by the model. Floor precision and floor F1 assess frame-level system-floor occupancy against the reference labels, whereas SS and SL F1 assess transition timing.

At 2,000 fine-tuning steps, teacher-forced validation loss decreases by approximately 30\%. \Cref{tab:moshi-results} summarizes the turn-taking results. In free generation, the fraction of reference turns taken rises from 0.44 to 0.85 and floor precision rises from 0.46 to 0.88. SS F1 at three-frame tolerance increases from 0.066 to 0.145, while SL F1 increases from 0.037 to 0.122. Median onset latency rises from 0.08 to 0.56 seconds because the fine-tuned model attempts nearly twice as many turns, including harder ones; the pretrained latency is conditioned on the minority of easy turns it detects.

\begin{table}[t]
\caption{Moshi turn-taking performance before and after 2,000 fine-tuning steps.}
\label{tab:moshi-results}
\centering
\small
\resizebox{\columnwidth}{!}{%
\begin{tabular}{llrr}
\hline
\textbf{Mode} & \textbf{Metric} & \textbf{Pretrained} & \textbf{Fine-tuned} \\
\hline
Free & Turns taken & 0.44 & 0.85 \\
Free & Floor precision & 0.46 & 0.88 \\
Free & SS F1, $\pm3$ frames & 0.066 & 0.145 \\
Free & SL F1, $\pm3$ frames & 0.037 & 0.122 \\
Teacher & Floor F1 & 0.893 & 0.962 \\
Teacher & Exact SL F1 & 0.60 & 0.93 \\
\hline
\end{tabular}}
\end{table}

Teacher forcing separates turn-timing knowledge from autoregressive drift, the divergence that accumulates when a model conditions on its own generated output. With correct context, floor F1 improves from 0.893 to 0.962. The largest improvement is exact release timing: zero-tolerance SL F1 rises from 0.60 to 0.93. Moshi therefore learns the intended transition locations to within a frame even though free generation can diverge from the single reference schedule.

We use the 2,000-step checkpoint as the fine-tuned model for all reported results. It nearly doubles turn coverage in free generation and sharpens release timing under teacher forcing, providing external evidence that the synthetic corpus teaches transferable full-duplex behavior.
\section{Related Work}
\label{sec:related}

\begin{figure*}[t]
    \centering
    \includegraphics[width=\textwidth,clip]{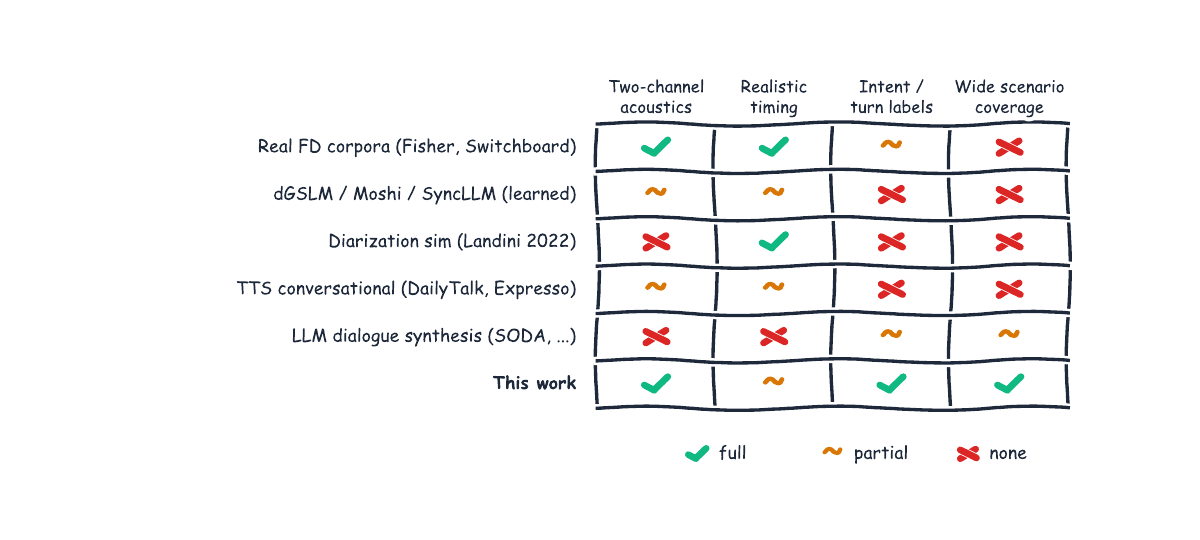}
    \caption{Comparison with approaches to conversational-speech data. Check marks denote full support, tildes partial support, and crosses no support. FD abbreviates full-duplex, and sim abbreviates simulation. Each existing approach covers a subset of the requirements targeted by this work.}
    \label{fig:related-comparison}
\end{figure*}

Our work connects full-duplex spoken dialogue, continuous turn prediction, conversational-audio simulation, and LLM dialogue synthesis. Each area supplies part of the training signal, but none combines two-channel acoustics, realistic timing, authored intent, and controlled coverage of floor-management phenomena (\Cref{fig:related-comparison}).

\subsection{Full-Duplex Dialogue and Turn Prediction}

Full-duplex models demonstrate the value of synchronized speaker streams. dGSLM learns from two-channel Fisher \cite{cieri2004fisher} audio and generates overlap, backchannels, and laughter without an explicit turn-taking module \cite{nguyen2023dgslm}. Moshi combines single-stream pretraining, recorded conversations, and synthetic dual-stream speech, but does not expose a controlled inventory of interruption intents \cite{defossez2024moshi}. SyncLLM interleaves fixed-duration speech-token chunks from both speakers, using synthetic speech for synchronization and Fisher \cite{cieri2004fisher} for overlapping behavior \cite{veluri2024syncllm}. These systems depend on uncontrolled conversations or synthesis without floor-intent labels.

State-based systems define the supervision target as a continuous decision. Freeze-Omni predicts whether to keep listening, interrupt, or finish without interruption, while SALMONN-omni interleaves audio streams with state-transition tokens at an 80-ms cadence \cite{wang2024freezeomni,tang2025salmonnomni}. Our labels derive each state from the authored event graph instead of inferring intent from voice activity.

Voice Activity Projection (VAP) predicts future joint voice activity from two-channel audio, capturing shifts, holds, and overlaps without manual labels \cite{ekstedt2022vap}. Earlier continuous models moved beyond endpoint detection, while TurnGPT showed that linguistic completion provides complementary turn-shift information \cite{skantze2017continuous,ekstedt2020turngpt}. Our approach preserves conversational intent, so similar overlaps receive different targets for successful interruptions, failed interruptions, and backchannels.

\subsection{Simulated Conversational Audio}

Speaker-diarization and overlapped-speech systems also arrange recorded segments on a shared timeline. Early mixtures sampled silences to reach a target overlap ratio. Landini et al. instead sample pause, gap, and overlap statistics measured from conversations \cite{landini2022simulated}. Their target is diarization: overlap types share one speaker-activity interpretation, source clips need not form a coherent dialogue, and the output is mixed rather than a clean human/system pair.

Our pipeline authors each event's purpose and relation to prior events, then uses TTS and forced alignment to determine duration and word landmarks. A barge-in, backchannel, and aside can therefore occupy similar activity regions but produce different system-side labels. Prior diarization simulation does not use aligned word boundaries to place authored events on a duplex timeline.

\subsection{Dialogue Synthesis and Timing}

LLM synthesis methods such as SODA, Dialogic, PLACES, UltraChat, and Baize generate multi-turn text for dialogue training \cite{kim2023soda,li2022dialogic,chen2023places,ding2023ultrachat,xu2023baize}. They specify what is said but omit acoustic duration, overlap, and floor-transfer timing. Conversational TTS datasets such as DailyTalk and Expresso add speech and prosody \cite{lee2023dailytalk,nguyen2023expresso} but remain sequential and lack intent-resolved labels. We instead author relational events and resolve their timing after synthesis.

Conversation analysis supplies timing constraints. Turn-taking theory describes a coordinated system that limits gaps and overlaps \cite{sacks1974systematics}. Cross-linguistic measurements show short response offsets, while speech-planning time implies that participants predict turn completion \cite{stivers2009universals,levinson2015timing}. Heldner and Edlund define the signed floor-transfer offset and show that transfer and within-speaker pause durations overlap \cite{heldner2010pauses}. We use these distributions as fallback priors, retain authored gaps, and measure landmarks inside rendered speech.

\subsection{Turn-Taking Benchmarks and Remaining Gap}

Full-duplex benchmarks evaluate pause handling, backchannel timing, interruptions, response latency, side-talk, and background speech \cite{lin2025fullduplexbench,lin2025fullduplexbench15}. These categories match our scenario registry, but the benchmarks probe deployed models rather than provide frame-level training data. Real two-channel corpora supply natural acoustics but cannot balance behaviors or identify the intent behind each overlap.

\Cref{fig:related-comparison} summarizes the gap. Our pipeline creates two-channel speech, combines measured speech timing with empirical silence distributions, retains intent for ambiguous activity patterns, and balances 42 phenomena across eight families.

\section{Summary}
\label{sec:summary}

The results support the central claim that authored conversational events can produce useful turn-taking supervision. The generation ablations show that the three diversity mechanisms affect different corpus properties. The structural hub prevents collapse to a turn-plus-backchannel pattern and restores interruptions, truncations, and varied floor-action sequences. The semantic hub adds meaning diversity beyond topic seeding, although the gain is limited by the strong distinct-topic baseline. Probability verbalization produces the clearest gains in wording and meaning diversity, while its effect on floor mechanics and timing is mixed. These differences show why lexical, semantic, and structural diversity must be measured separately.

The internal semantic-VAD results show that the authored labels form a learnable causal prediction target. The model identifies the sparse start-speaking and start-listening boundaries and remains stable under user-channel noise. The finer label space remains learnable but exposes the cost of rare classes and distinctions that are weakly expressed in the waveform. Merging labels that distinguish user-originated from companion-originated backchannels and interruptions improves performance under both acoustic conditions, indicating that the taxonomy should retain distinctions supported by audio rather than every distinction available in the authored schema.

Moshi provides evidence that the corpus transfers beyond the internal classifier. Fine-tuning causes the model to take more appropriate turns, improves floor precision, and sharpens release timing when evaluated with the correct conversational context. Free-generation boundary scores remain lower because autoregressive speech can diverge from the single reference timeline, but the combined gains show that the model learns the intended floor transitions. Together, these results establish controlled, intent-labeled synthesis as a practical complement to uncontrolled conversational recordings for training full-duplex dialogue systems.

Future work should improve the multilingual scenario prototypes for live translation and language teaching. The event representation already supports per-event languages, concurrent interpretation, multiple timing references, and interpreter lag (the delay between source speech and its translation), but the generated samples do not yet meet the quality standard of the released corpus. Adding these scenarios will require better bilingual coherence, translation fidelity, and control of cross-language timing and speech realization. Other priorities identified during corpus analysis are adding competing background talkers, increasing held-floor and within-turn-pause coverage to correct the bias toward completed turns, and extending the language coverage of the pipeline beyond English and Mandarin.

\begingroup
\renewcommand{\footnotesize}{\fontsize{9}{11}\selectfont}
\bibliographystyle{IEEEtran}
\bibliography{citations}
\endgroup

\clearpage
\appendices
\section{Conversation-Authoring Prompts}
\label{sec:appendix}

The generation script constructs its prompts from the templates below. Angle-bracketed text marks values inserted at runtime from the merged metadata, scenario registry, canonical example, diversity hubs, topic seed, or validator. The metadata block includes the event-field definitions, timing policy, reaction-time ranges, multilingual guidance, and other authoring rules. In the templates, \texttt{MS\_POLICY} denotes the millisecond timing policy, \texttt{pause\_affinity} controls optional pacing guidance, and an anchor is a referenced event or word boundary used to place or truncate speech. The field \texttt{gap\_before\_ms} specifies the signed gap before an event; \texttt{cue.ref} identifies the referenced event, \texttt{cue.ref\_word} names a word anchor, and \texttt{offset\_ms} specifies a timing displacement in milliseconds. For an interruption, \texttt{truncate.until\_word} names a word in the interrupting utterance used to time the cut to the interrupted utterance. Prompt wording is retained from the source templates rather than copyedited.

\begin{promptbox}[colback=blue!4,colframe=blue!35!black]{System prompt: conversation author}
You are an expert author of two-channel (Human<->AI) conversational speech scenarios in the v<schema_version> event-list schema. You write the SCRIPT: who speaks, what they say, and how each turn attaches to an earlier one via a cue. You NEVER use absolute timestamps---placement is expressed only through cues and gap_before_ms (which you will author with consideration to the natural rhythm and pacing of the conversation).

Channels/voices: user & companion are the Human side (channel 1); system is the AI (channel 2). Use 'companion' only for side-conversations (asides).

<EVENT FIELDS, TIMING, MS_POLICY, REACTION-TIME BUCKETS,
 LIVE-TRANSLATION INSTRUCTIONS, AND RULES FROM METADATA>

OUTPUT CONTRACT: respond with ONE JSON object and nothing else (no markdown, no prose). Shape:
{"conv_desc":"<one sentence naming the NEW topic/context>",
 "events":[{"id":"e0","voice":"user","act":"turn",
 "cue":{"ref":null,"point":"start"},"text":"...","lang":"tag"}, ...]}

Conversations might be multilingual, so tag each event appropriately.
conv_desc states WHAT the new conversation is about (its topic/situation)---NOT the phenomenon, which is fixed by the scenario and given below.

Hard requirements you MUST satisfy (they are validated automatically):
  - ids are e0..eN, unique; every cue.ref points to an EARLIER event (null only on e0).
  - any ref_word/until_word MUST appear verbatim in the referenced/own text.
  - barge_in events set barge_type and an offset_ms from the matching bucket, plus a truncate.until_word taken from the barge's OWN text; a failed barge sets interrupts:false and no truncate.
  - same-channel turns must not overlap; keep the conversation ~30-60 s when spoken.

<LANGUAGE-SPECIFIC AUTHORING GUIDANCE>
\end{promptbox}

\begin{promptbox}[colback=green!4,colframe=green!35!black]{User prompt: scenario re-authoring}
Scenario to re-author: <scenarioID>---<scenario_name>
(family <family>: <family_name>).

PHENOMENON to preserve (fixed for this scenarioID---your script MUST match the phenomenon): <scenario_desc>

Here is a generic example conversation for this scenario (its topic is in conv_desc). Study its acts/cues, then write a DIFFERENT script on a NEW topic---new situation, new wording, new entities, new chronemics, new turn structure---that demonstrates the SAME phenomenon just as clearly, and name that new topic in conv_desc:

<CANONICAL EXAMPLE AS FORMATTED JSON>

<OPTIONAL PACING GUIDANCE FROM pause_affinity>

<OPTIONAL SEMANTIC-HUB EXAMPLES, WITH AN INSTRUCTION TO VARY
 SITUATION, TOPIC/MEANING, VOCABULARY, AND ENTITIES>

<OPTIONAL STRUCTURAL-HUB FLOOR FINGERPRINTS, WITH AN
 INSTRUCTION TO VARY ACT ORDER, OVERLAP WORDS, AND TIMING>

<OPTIONAL SOFT TOPIC SEED AND ITS VOCABULARY>

Write the new event list now as the single JSON object specified in the system message.
\end{promptbox}

\begin{promptbox}[colback=cyan!4,colframe=cyan!40!black]{Semantic-hub prompt block}
You have also ALREADY written these conversation(s) for this scenario---make this one genuinely DIFFERENT in SITUATION, TOPIC/MEANING, VOCABULARY, and ENTITIES (not a paraphrase, and not the same subject in other words):

<SEMANTIC-CORESET EXAMPLES AS COMPACT JSON>

Each example contains its conv_desc and, for every event, only the voice, act, and text. When structural-hub examples are also present, the prompt adds:

Their turn-taking and timing are handled separately below, so here focus on the content.
\end{promptbox}

\begin{promptbox}[colback=yellow!7,colframe=yellow!45!black]{Structural-hub prompt block}
Here are the FLOOR MECHANICS you have already produced for this scenario, shown as the ordered act sequence (these are the structural CORNERS of what you've written---some may be the same conversations shown in full above, here reduced to just their turn-taking and timing):

  sample 1: <ordered floor fingerprint>
  sample 2: <ordered floor fingerprint>
  ...

Make THIS sample's floor mechanics noticeably DIFFERENT from the samples in THIS list: use a different act-sequence, different barge-in opener words and different backchannel tokens, and choose different---but still legal---timing numbers. Vary each barge-in's reaction offset_ms within its barge_type range, vary the gap_before_ms pauses/overlaps between turns, the number of turns, and how much speech overlaps. Do NOT reuse the same numbers or the same interruption/backchannel words as this list.
\end{promptbox}

\begin{promptbox}[colback=violet!4,colframe=violet!35!black]{Probability-verbalized batch variant}
OUTPUT CONTRACT: generate 5 DISTINCT conversation instances for this scenario. Respond with ONE JSON object and nothing else (no markdown, no prose). Shape:

{"responses":[
  {"conv_desc":"<one sentence naming the NEW topic/context>",
   "events":[<full event list>],
   "probability":<float 0.0-1.0>}, ...]}

Each response MUST include its new conv_desc, full event list, and your estimated probability of that response given the input prompt, relative to the full distribution of responses you could give.

Write 5 new, DISTINCT conversations. They all address the phenomenon and topic above---make them differ from EACH OTHER in situation, wording, entities, and chronemics.
\end{promptbox}

\begin{promptbox}[colback=red!5,colframe=red!55!black]{Validation-retry prompt}
Your previous JSON (the assistant turn just above) FAILED these checks---return a CORRECTED full JSON object (same output contract) that fixes ALL of them while keeping everything that was already correct, and keep every cue self-consistent with your text:

  - <validator error 1>
  - <validator error 2>
  - ...
\end{promptbox}

\subsection{Scenario-Judge Prompts}
\label{sec:judge-prompts}

The judge reports problems without rewriting the conversation. Angle-bracketed text denotes runtime inputs; the anchoring block is included only when the anchor inventory is nonempty. The Chinese filler in the system prompt is romanized as \textit{en} for typesetting. An \texttt{interpret} chunk is a segment of spoken translation; its ear--voice span is the delay between the source speech and the translated speech. The source user template says \emph{two axes}, although the system template lists three: semantics, anchoring choice, and formatting.

\begin{promptbox}[colback=orange!4,colframe=orange!45!black,listing options app={escapeinside={(*}{*)}}]{System prompt: scenario judge}
You are a STRICT reviewer of two-channel (Human<->AI) conversational-speech scenarios written in an event-list schema. You do NOT rewrite anything---you REPORT problems as JSON. Structure, the cue reference graph, whether an anchor word is present in the text, and timing legality have ALREADY been validated by separate deterministic tooling; do NOT re-report those. Judge ONLY these factors:
  1. SEMANTICS: is every turn a coherent, on-topic, sensible response to what came before it? Flag contradictions, non-sequiturs, a system turn that ignores or misunderstands the user, or garbled/meaningless text.
  2. ANCHORING CHOICE: every event attaches to one or more EARLIER events (cue.ref, or cue.refs for concurrent speech) at a point (end / word_start / word_end), and a barge_in / backchannel / cue may further pin to a specific word (cue.ref_word) or truncate at one (truncate.until_word). The reference graph is legal and every anchor word EXISTS (already validated); you judge whether the CHOICES are the right ones to produce the intended timing/overlap:
     You are reviewing the FINAL scenario AFTER all timing scaffolding has been sampled, so a numeric cue.offset_ms is EXPECTED on both barge_in events (reaction time) and interpret chunks (ear-voice span). NEVER flag the mere PRESENCE of offset_ms, and never claim it 'should be automatic / omitted / removed'---that guidance is for the AUTHOR, not for this finished artifact. The only offset you may judge is whether a BARGE_IN's offset VALUE fits the intended interruption timing.
     - WRONG EVENT: an event anchored to the wrong earlier event---e.g. a reply whose cue.ref points at the wrong predecessor, a barge_in interrupting the wrong turn, or an interpret chunk whose source ref is not the line it translates. This applies even when there is NO ref_word---the event->event edge itself can be wrong.
     - WRONG WORD: a plausible-but-wrong trigger word (e.g. reacting to a word that isn't the one being corrected), or a word that sits on the wrong event for the intended reaction.
     - WRONG INSTANCE: when the trigger word repeats in the referenced text, the occurrence index selects a semantically-wrong repetition (e.g. reacting to the first 'Thursday' when the correction is about the second). (Do NOT report a merely invalid occurrence value like 0---that is mechanical and already validated.)
     An interpret act's cue.offset_ms is the pipeline-sampled ear-voice span (seconds-scale, grows across the turn)---it is expected and correct; do NOT flag its presence OR its magnitude (it is NOT a barge_in reaction time). Judging an interpret chunk's SOURCE anchor (its ref / ref_word) is still in scope.
  3. FORMATTING: small field-value mistakes that will break the render even though the structure is legal. Specifically: an event's `lang` tag MUST be one of 'en'/'eng'/'englsih' for English or 'zh'/'zho'/'cmn'/'chinese'/'mandarin'/'zh-cn'/'zh-tw'/'zh-hans'/'zh-hant' for Chinese (this corpus is English + Chinese only)---flag any other value, since the TTS accepts only English/Chinese and a stray tag crashes the render. Keep this axis NARROW: only the lang field---do NOT invent other formatting nits.

Be conservative: flag only GENUINE problems a careful human would fix. If the scenario is coherent and the anchors are sensible, return ok=true with an empty errors list.

<RULE SUMMARY FROM summarize_rules(meta)>

CRITICAL---your fix_hint suggestions MUST themselves obey every rule above, because a fix is fed back to the author verbatim and re-validated. In particular:
  - Write ALL numbers, money, times, and symbols as the SPOKEN WORDS, never as digits or symbols: 'four hundred' not '400', 'forty three dollars' not '$43', 'eight o'clock' not '8:00', 'nineteen eighty eight' not '1988', 'fifteen percent' not '15%'. A fix_hint containing a digit or a symbol like $ % (*$^\circ$*) : is itself INVALID and will be rejected by the deterministic validator.
  - Never suggest anchoring a cue.ref_word / truncate.until_word on a filler or non-lexical token (um, uh, mm-hmm, oh, (*\textit{en}*) ...)---the forced aligner emits no token for it. Suggest a CONTENT word.
  - occurrence is 1-indexed (>= 1). Any word you name as a new anchor must actually appear in the relevant event's text.

OUTPUT CONTRACT: respond with ONE JSON object and nothing else (no markdown, no prose). Shape:
{"ok": true|false, "errors": [{"event_id": "e1", "axis": "semantics|anchoring|formatting", "problem": "<what is wrong>", "fix_hint": "<concrete change that would fix it>"}]}
ok MUST be false if and only if errors is non-empty.
\end{promptbox}

\begin{promptbox}[colback=green!4,colframe=green!35!black]{User prompt: scenario judge}
Scenario <scenarioID>---<scenario_name> (family <family>).
PHENOMENON this scenario must portray (fixed): <scenario_desc>
TOPIC of this specific conversation: <conv_desc>

EVENTS (in authoring order):
<EVENTS AS FORMATTED JSON, RETAINING id, voice, act, lang,
 cue, truncate, barge_type, AND text>

ANCHORING GRAPH to check (right EVENT to attach to, right WORD, right INSTANCE):
<ANCHOR INVENTORY: ATTACH EDGES AND WORD ANCHORS WITH
 THEIR POINTS AND OCCURRENCE INDICES>

Review the two axes and return the JSON verdict specified in the system message.
\end{promptbox}

\clearpage
\section{Labeling Examples}
\label{sec:label-examples}

The following examples show how the same two-channel waveform can be represented at three levels of label granularity. The core-4 view retains only the four basic speaking and listening actions. The granular view adds distinctions such as backchannels, interruptions, and types of silence. The granular-all view additionally distinguishes contested-floor speech, system pauses while holding the floor, and whether human-channel speech comes from the user or a companion. These are visualization label sets; granular-all is not the 13-token \texttt{all\_merged} benchmark space, which merges the user/companion variants of yield and hold-backchannel.

\begin{figure}[!htbp]
    \centering
    \includegraphics[width=\columnwidth]{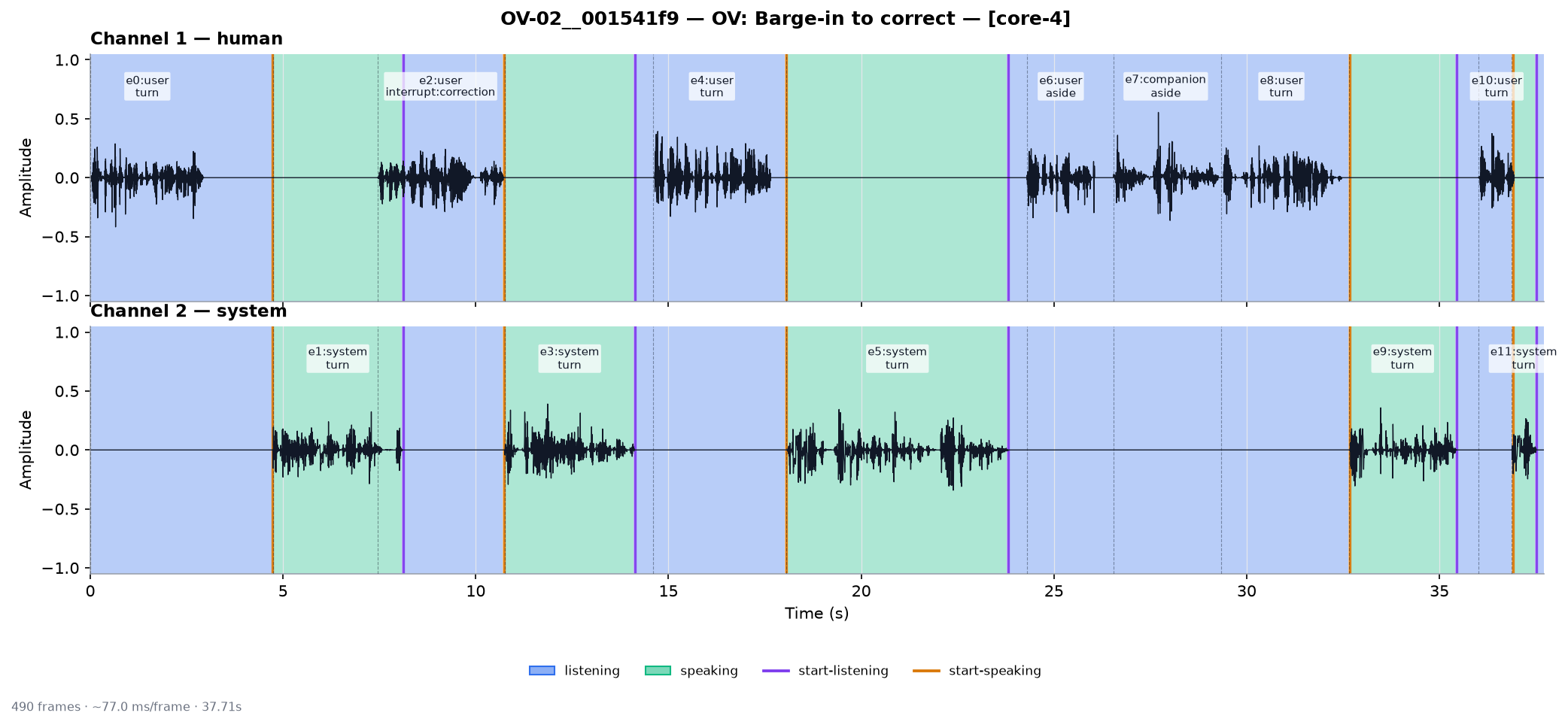}
    \par\smallskip {\small (a) Core-4 labels.}\par\medskip
    \includegraphics[width=\columnwidth]{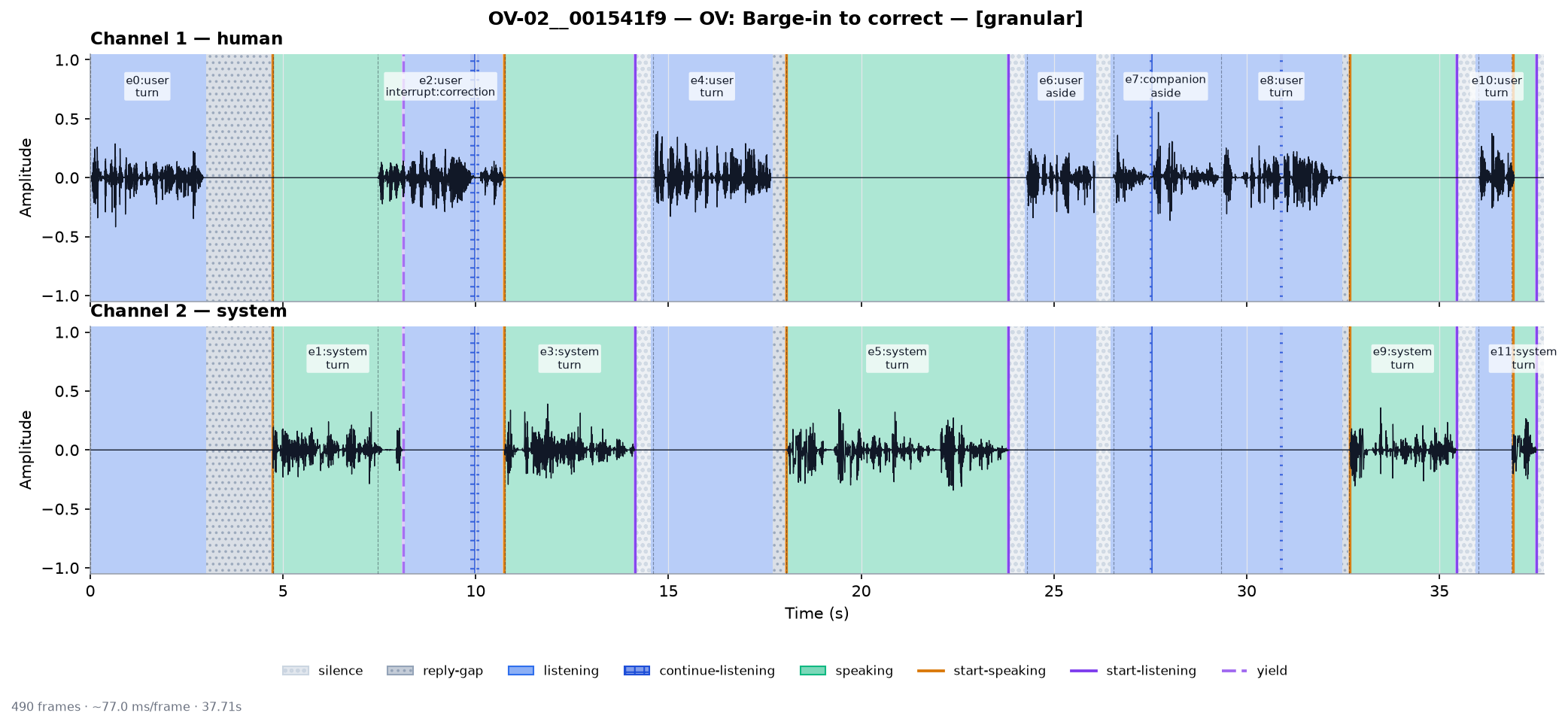}
    \par\smallskip {\small (b) Granular labels.}\par\medskip
    \includegraphics[width=\columnwidth]{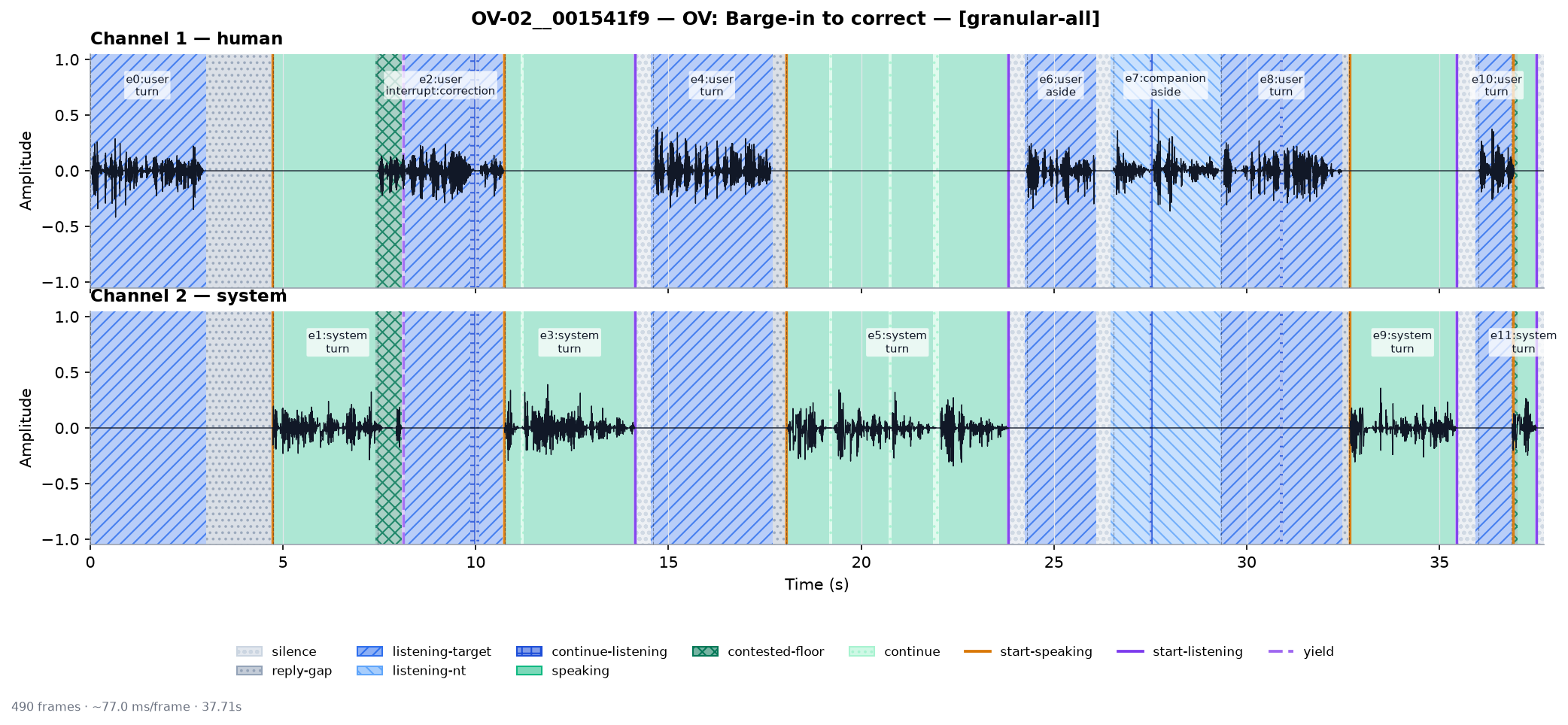}
    \par\smallskip {\small (c) Granular-all labels.}\par
    \caption{Three label granularities for OV-02, a correction barge-in. Each panel shows the same human and system waveforms while increasing the detail of the system-side label taxonomy.}
    \label{fig:label-examples-ov02}
\end{figure}

\begin{figure}[!htbp]
    \centering
    \includegraphics[width=\columnwidth]{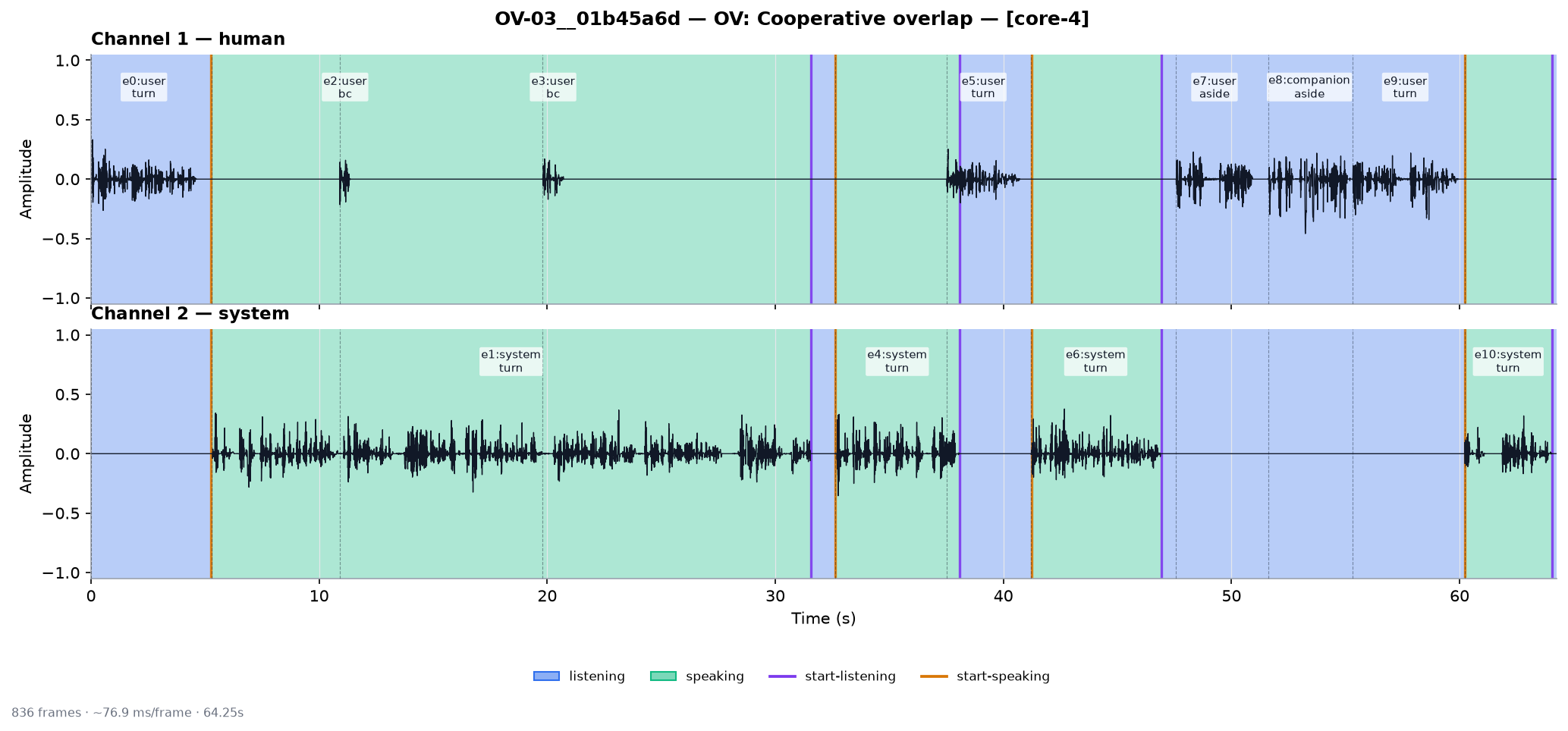}
    \par\smallskip {\small (a) Core-4 labels.}\par\medskip
    \includegraphics[width=\columnwidth]{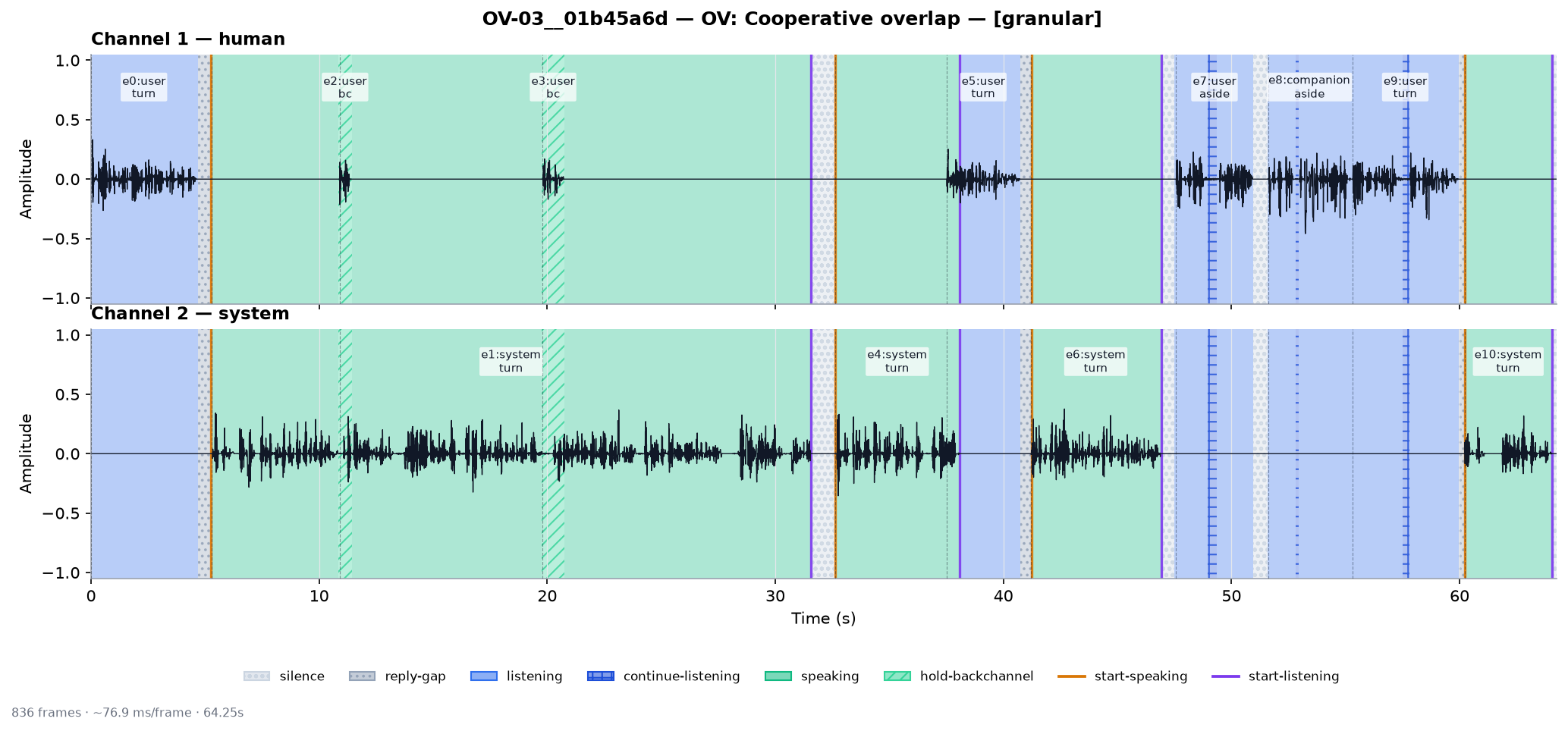}
    \par\smallskip {\small (b) Granular labels.}\par\medskip
    \includegraphics[width=\columnwidth]{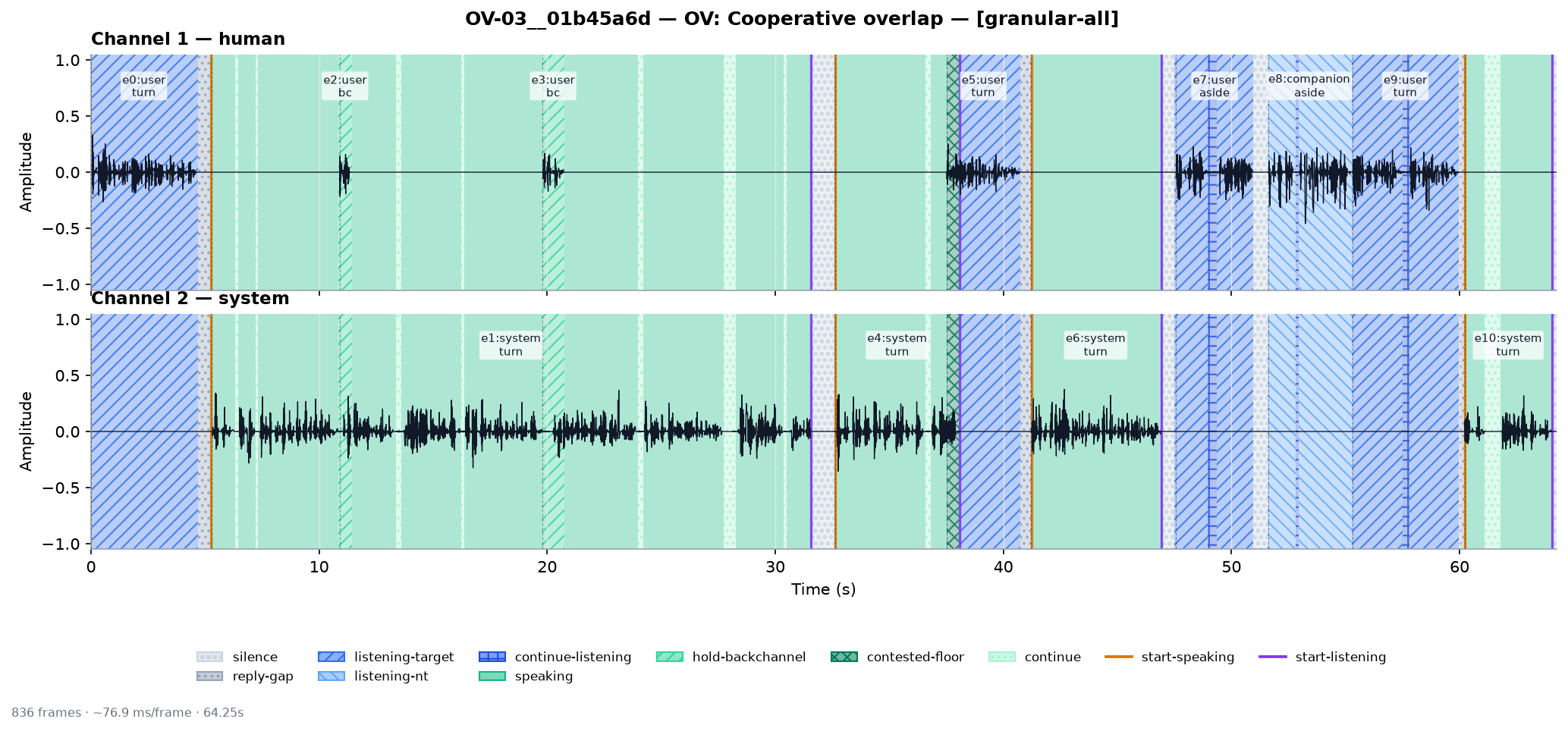}
    \par\smallskip {\small (c) Granular-all labels.}\par
    \caption{Three label granularities for OV-03, a cooperative-overlap conversation with backchannels. The finer views distinguish additional floor states that collapse into the four core actions.}
    \label{fig:label-examples-ov03}
\end{figure}

\end{document}